\documentclass[aps,floats,nofootinbib,preprint, superscriptaddress]{revtex4}

\usepackage{epsfig}
\usepackage{amssymb}
\usepackage{bm}
\usepackage{hyperref}
\usepackage{color}
\usepackage{amsmath}

\newcommand{\be}{\begin{equation}}
\newcommand{\ee}{\end{equation}}
\newcommand{\bea}{\begin{eqnarray}}
\newcommand{\eea}{\end{eqnarray}}

\renewcommand{\slash}{\displaystyle{\not}}

\newcommand{\nn}{\nonumber}

\newcommand{\sci}[2]{#1$\times$10$^{\text{#2}}$}

\newcommand {\ga} {\ {\raise-.5ex\hbox{$\buildrel>\over\sim$}}\ }
\newcommand {\la} {\ {\raise-.5ex\hbox{$\buildrel<\over\sim$}}\ }

\begin{document}

\title{Anapole Dark Matter at the LHC}
\author{Yu Gao} \email{yugao@physics.tamu.edu}
\affiliation{Department of Physics, University of Oregon, Eugene, OR 97403, USA}
\affiliation{Department of Physics and Astronomy, Texas A \& M University, College Station, TX 77843, USA}
\author{Chiu Man Ho} \email{cmho@msu.edu}
  \affiliation{Department of
  Physics and Astronomy, Vanderbilt University, Nashville, TN 37235, USA}
   \affiliation{Department of
  Physics and Astronomy, Michigan State University, East Lansing, MI 48824, USA}
\author{Robert J. Scherrer} \email{robert.scherrer@vanderbilt.edu}
  \affiliation{Department of
  Physics and Astronomy, Vanderbilt University, Nashville, TN 37235, USA}
\date{\today}

\begin{abstract}

The anapole moment is the only allowed electromagnetic moment for Majorana fermions. Fermionic dark matter acquiring an anapole
can have a standard thermal history and be consistent with current direct detection experiments. In this paper, we calculate the collider
monojet signatures of anapole dark matter and show that the current LHC results exclude anapole dark matter with mass less than
100 GeV, for an anapole coupling that leads to the correct thermal relic abundance.

\end{abstract}
\maketitle

\section{Introduction}

The nature of the dark matter that constitutes most of the
nonrelativistic density in the universe
remains unresolved.  While the leading candidates are usually considered to be either a massive particle interacting via the weak force (WIMP),
or an axion (see, e.g., the recent review in Ref.~\cite{axion}), there has been a great deal of recent interest in the possibility that the
dark matter interacts electromagnetically.  Dark matter with an integer electric charge number $\sim {\cal O}(1) $ has long been ruled out, and
even millicharged dark matter is strongly disfavored~\cite{Steigman}.  Hence, the most attention has been paid
to models in which the dark matter particle has an electric or magnetic dipole moment, which we will call generically
dipole dark matter (DDM)~\cite{Pospelov,Sigurdson,Gardner,Masso,Fitzpatrick,Cho,Heo1,Heo2,Banks,Barger1,Fortin,
Nobile,Barger2,Heo3}. If one assumes a thermal production history for the dark matter, fixing the dipole moment coupling to provide the correct
relic abundance, then the corresponding rate in direct detection experiments rules out a wide range of DDM
mass \cite{Sigurdson,Masso,Fortin}.

An alternative to DDM is a particle with an anapole moment.  The idea of the anapole moment was first proposed by
Zel'dovich \cite{Zeldovich} and mentioned in the context of dark matter by Pospelov and ter Veldhuis \cite{Pospelov}.
More recently, the properties of anapole dark matter (ADM) were investigated in detail by Ho and Scherrer \cite{Ho:2012bg}.
(See also the model of Fitzpatrick and Zurek \cite{Zurek}, in which the anapole couples to a dark photon rather than
a standard-model photon).
Anapole dark matter has several advantages over DDM.  The anapole moment is the only allowed electromagnetic moment
if the dark matter is Majorana, rather than Dirac.  The annihilation is exclusively $p$-wave, and the anapole
moment required to give the correct relic abundance produces a scattering rate in direct detection experiments that
lies below the currently excluded region for all dark matter masses (although see our discussion of LUX in Sec. V).

Here we extend the discussion of Ref. \cite{Ho:2012bg} to consider collider signatures of anapole dark matter.
As we show in the next section, the anapole Lagrangian allows for the pair production of anapole dark matter, along with a
jet that makes the event visible. The dark matter is then manifested as missing energy + monojet \cite{Monojet}.
Mono-photon \cite{Mono-photon}, mono-Z \cite{Mono-Z} or mono-Higgs \cite{Mono-Higgs} signals are subdominant in our model as their cross-sections suffer from smaller couplings.\footnote{One recent study in \cite{HeavyMajorana} investigated similar collider signals for a model with a heavy Majorana neutrino being the dark matter.} Further, these other final states are produced only in $q \,\bar q$
interactions, and the $\bar{q}$ at the LHC is a sea quark. In comparison,
the monojet event can be produced from a $q\,g$ initial state which is not
suppressed by proton's parton distribution. We use the latest LHC monojet results to calculate the corresponding limits on the anapole moment in Sec. III.  In Sec. IV, we extend the thermal abundance calculations of Ref. \cite{Ho:2012bg} up
to higher dark matter particle masses ($\sim$ 1 TeV) and show that the dominant annihilation channel is
$\chi \,\chi \rightarrow W^+W^-$ when $m_\chi > m_W$.  Our results are discussed in Sec. V.  We find that
$m_\chi < 100$ GeV is excluded by the LHC.

\section{Anapole Dark Matter Monojet}

\begin{figure}
\includegraphics[scale=0.7]{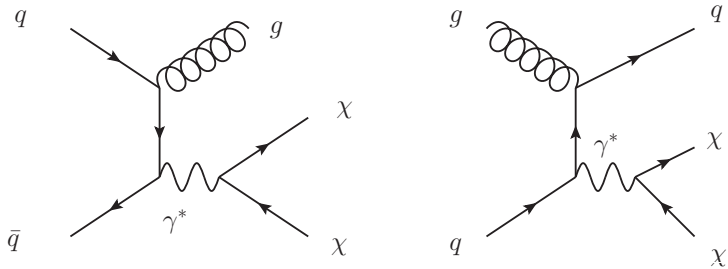}
\caption{Monojet diagrams at colliders. Diagrams with permuted initial particles also contribute.}
\label{fig:diagrams}
\end{figure}

We denote the anapole Lagrangian as
\be
\label{Lagrangian}
{\cal L}_{int} = g_A \,\bar{\chi}\, \gamma^\mu \,\gamma^5\, \chi \,\partial^\nu F_{\mu\nu}\,,
\ee
where $g_A\equiv g/\Lambda^2$, with $g$ being the coupling constant
and $\Lambda$ the cutoff scale, as in Ref. \cite{Ho:2012bg}. In Eq. (\ref{Lagrangian}), $\chi$ is the ADM
particle, which we take to be a Majorana fermion.

The leading Feynman diagrams for collider monojets are illustrated in Fig.~\ref{fig:diagrams}.
The only visible part of the event is the single jet that recoils against the $\chi\chi$ pair. In the center of mass (COM) frame, the invariant mass of the ADM pair is related to the jet energy as
\be
M^2_{\chi\chi}=M^2_{\gamma^*}= s-2\,\sqrt{s} \,E_{j}\,,
\label{eq:invmass}
\ee
where $\sqrt{s}$ is the total event energy and $E_{j}$ is the jet energy. For a relatively soft jet $(E_j \ll \sqrt{s})$, the virtual photon
mass is $\sqrt{s}$ and the anapole coupling is not suppressed.  As the low $E_{j}$ contribution
accounts for the majority of cross-section due to infrared
divergence, the $p_T$ and energy cuts determine the signal event rate. This leads to the effect that at low $m_\chi$, the cross section becomes
insensitive to $m_\chi$. It is worth noting that the virtual photon propagator also has a pole at $E_{j}=\sqrt{s}/2$, but it is cancelled by the fact that the anapole vertex vanishes at a physical photon.

The squared amplitude for the first diagram in Fig.~\ref{fig:diagrams} is
\be
|M|^2 = \frac{512 \,\pi^2\, g_A^2\, \alpha_s \,\alpha \cdot\, Q^2}{P_{T,j}^2}\,
E_0^4 \,\cdot\, {\cal K}(E_j,\,E_\chi,\theta_j,\,\theta_\chi)\,,
\label{eq:Melement1}
\ee
where $E_0$ is the beam energy in the center of mass frame, $\alpha = 1/137$ is the fine structure constant,
$E_j$, \,$E_\chi$,\, $\theta_j$ and $\theta_\chi$ are respectively
the energy and scattering angles of the radiated jet and one of the ADM,
$Q$ is the electric charge of the relevant quark, and $P_{T,j}$ in the denominator is the transverse momentum of the single jet. For the second diagram, we have
\be
|M|^2 = \frac{128\, \pi^2\, g_A^2 \,\alpha_s\,\alpha \cdot Q^2}{(\,1+\cos\theta_j\,) E_j}
\,E_0^3 \cdot {\cal K'}(\,E_j\,,E_\chi,\,\theta_j,\,\theta_\chi\,)\,.
\label{eq:Melement2}
\ee
The kinematic factors ${\cal K},\,{\cal K'}$ are given in Appendix~\ref{app:msq}.
Both diagrams are sensitive to jet
$p_T$ and jet energy. The $\chi$ mass is irrelevant unless it starts to suppress the phase space at the TeV scale.

Now consider the kinematics of these events.
The Feynman rule for the $\chi \chi \gamma$ vertex reads
\be
2\,\slash{p}_{\gamma^*}\,\gamma^5\, p^\mu_{\gamma^*} - 2\,\gamma^\mu\,\gamma^5\cdot p^2_{\gamma^*}\,,
\label{eq:feynmanrule}
\ee
where $p_{\gamma^*}$ is the 4-momentum of the off-shell photon and the factor of 2 comes from $\chi$ being self-conjugate. By Eq.~(\ref{eq:invmass}), the effective coupling at this vertex grows as the event energy squared. This leads to a rather stringent constraint from the LHC.

Combining Eq.~(\ref{eq:invmass}) with Eq.~(\ref{eq:Melement2}), which corresponds to
the dominant diagram in LHC monojet searches, we see that at any given center of mass energy, the cross-section
is maximized at low $E_j$ and large $M_{\gamma^*}$. Namely, the final state jet is favored to sit at the lowest jet $E_T$ that passes
the event selection, and the $\chi \chi$ pair takes up the bulk of the energy (as missing energy). The cross-section increases
quickly with the center of mass energy until it becomes suppressed by the parton momentum distribution (PDF) in protons. This undesirable high-energy behavior arises from the high-dimensionality of our effective anapole operator. At this point, it is worth checking the energy flowing into the $\chi\chi\gamma^*$ vertex:
\be
M_{\gamma^*}\,\approx\, \sqrt{s}\,,
\ee
such that the effective operator would remain a good approximation with cut-off scale $\Lambda$ above the event energy. Notably with the LHC running at multiple TeV,
corrections to the effective operator should emerge when the event energy comes close to $\Lambda$.

\section{Collider Bounds}

To compute the collider constraints
at the LHC, we implemented the anapole Lagrangian in the {\it Calchep} package and calculated the signal rates at the parton level. We use CTEQ6L~\cite{CTEQ6} for the proton PDF.  
For the one-jet cross-section at the parton level, a K-factor would be expected; we expect this correction to be no more than one order of magnitude,
and it does not qualitatively alter our results.

\begin{figure}[h!]
\includegraphics[scale=0.9]{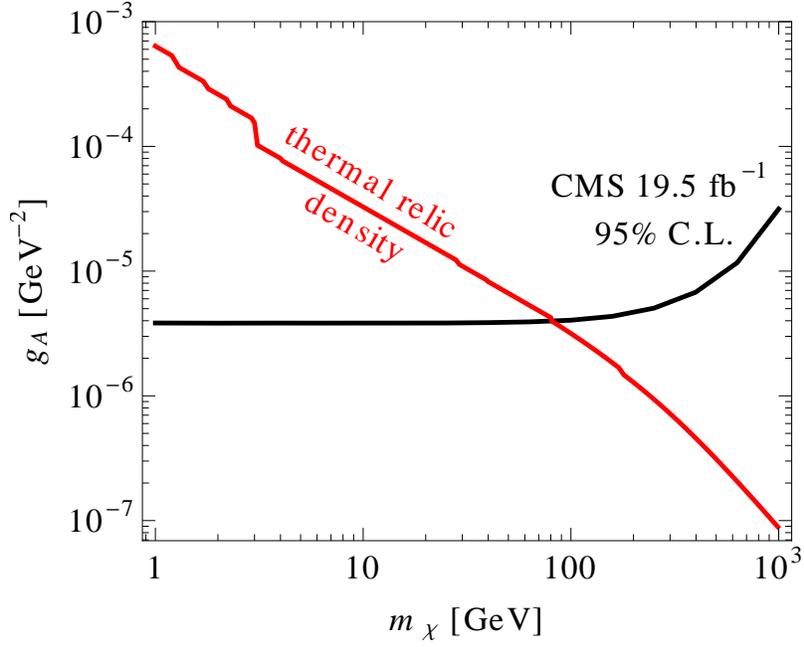}
\caption{Black curve gives the upper bound on the anapole moment, $g_A$, from CMS monojet data for
the indicated anapole dark matter mass, $m_\chi$.  Red curve gives the value of $g_A$ needed to
produce the observed dark matter abundance as a function of $m_\chi$.}
\label{fig:bounds}
\end{figure}

Due to collinear and infrared divergences, significant $E_{j}$ and  jet pseudo-rapidity $\eta_j$ cuts must be applied. In Table~\ref{tab:data}, we list the kinematic cuts and observed data from the latest LHC results. The experimental cuts include combinations of $E_j$, $p_T$, missing transverse energy (MET) and jet number ($N_j = 1$) bounds. For our anapole calculation, only the event $p_T$ and jet energy cuts are relevant. In Table~\ref{tab:data}, we show the choice of jet $p_T$ cut that optimizes the constraint on the effective coupling $g_A$ in the low ADM mass limit. Both CMS and ATLAS present experimental results in multiple sets of cuts. Here we only show the cuts that give the most stringent constraint. In Fig.~\ref{fig:bounds}, we illustrate CMS's monojet constraint on $g_A$ in combination with the value of $g_A$ that yields the correct thermal relic abundance. Note that the collider constraint at small ADM mass is rather stringent.

\begin{table}[h!]
\begin{tabular}{c|c|c}
\hline
Experiment ~~ &~~  Monojet cuts ~~ & ~~  allowed $g_A^*$\\
\hline
CMS 8 TeV, 19.5 fb$^{-1}$~\cite{cms2013:8TeV} ~~ & ~~  $\slash{E}_T > 450$ GeV, $|\eta_j|<$2.4 ~~ & ~~\sci{4}{-6} @ 95\% C.L.\\
ATLAS 8 TeV, 10.5 fb$^{-1}$~\cite{atlas2013:8TeV} ~~ & ~~ $\slash{E}_T > 220$ GeV, $|\eta_j|<$2.0 ~~ 	&	~~\sci{6}{-6} @ 95\% C.L.	\\
\hline
\multicolumn{3}{l}{$*$~ in the low $m_\chi$ limit.}
\end{tabular}
\caption{LHC monojet data and upper-bound on effective anapole coupling $g_A$.}
\label{tab:data}
\end{table}



\section{Improved Relic Abundance Calculation}

In Ref. \cite{Ho:2012bg}, the thermal relic abundance of the anapole dark matter particle was calculated, 
where $m_\chi$ was extended up to 80 GeV and only the annihilation into light species was considered. In this paper, we are interested in
ADM masses as large as $1$ TeV, so two additional annihilation channels open up: $\chi\chi\rightarrow W^+W^-$ and $~t\bar{t}$.

\begin{figure}[h]
\includegraphics[scale=0.5]{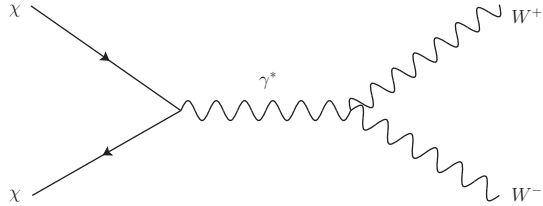}
\caption{$\chi\,{\chi} \rightarrow W^+\,W^-$.}
\label{fig:WW}
\end{figure}


The Feynman diagram for $\chi \chi\rightarrow W^+W^-$ is illustrated in Fig.~\ref{fig:WW}. Its squared amplitude is given as
\bea
|M|^2&=&128\,\pi \, g_A^2\, \alpha\, m_\chi^2\, \left\{
\,(\,1-y\,)\,[\,(\,4\,y+12+3\,y^{-1}\,) \right.\\ \nn
 &&\left. -(\,4\,y-4+3\,y^{-1}\,)\,\cos^2\theta \,]
\,\right\}\,p_\chi^{2}\,,
\eea
where $y\equiv m_W^2/m_\chi^2$. The nonrelativistic annihilation cross-section is
\bea
\label{sigW}
\sigma_{\chi \chi \rightarrow W^{+}W^{-}} \,v_{\textrm{rel}} = \frac{2}{3}\, g_A^2\, \alpha \,m_\chi^2 \,(\,1-y\,)^{\frac{3}{2}}\,(\,4\,y+20+3\,y^{-1}\,)\,v_{\textrm{rel}}^2\,,
\eea
where $v_{\textrm{rel}}$ is the relative velocity between the annihilating ADM particles.

As in Ref. \cite{Ho:2012bg}, we will make the simplifying assumption that annihilations into $W^+W^-$ are
negligible for $m_\chi < m_W$, and use the $m_\chi \gg m_W$ limit of Eq. (\ref{sigW}) for
the case where $m_\chi > m_W$, so that
\begin{equation}
\label{sigW2}
\sigma_{\chi \chi\rightarrow W^{+}W^{-}} \,v_{\textrm{rel}} = 2 \,g_A^2\, \alpha\, \frac{m_\chi^4}{m_W^2} \,v_{\textrm{rel}}^2\,.
\end{equation}
This will provide an accurate estimate of the relic abundance
as long as $m_\chi$ does not lie close to $m_W$. Using $\langle\, v_{\textrm{rel}}^2 \,\rangle = 6T/m_\chi$, we obtain
\begin{equation}
\label{sigW22}
\langle\,\sigma_{\chi \chi\rightarrow W^{+}W^{-}} \,v_{\textrm{rel}}\,\rangle = 12 \,g_A^2\, \alpha\, \frac{m_\chi^4}{m_W^2} \,\left(\frac{T}{m_\chi}\right)\,.
\end{equation}

The $\chi \chi\rightarrow t\bar{t}$ channel can be treated identically to other quark-antiquark annihilation channels considered in Ref.~\cite{Ho:2012bg}. Thus, the total annihilation cross-section of $\chi \chi$ into charged fermion-antifermion pairs $f\bar{f}$ is given by 
\footnote{Note, in this paper, we perform the calculations without a pre-factor of 1/2 in the anapole Lagrangian, so the annihilation cross-section formulas come with a coefficient of 4 in comparison to those in Ref.~\cite{Ho:2012bg}. 
}
\begin{equation}
\label{oldsig}
\sum_{m_f < m_\chi}\,\langle\,\sigma_{\chi\, \chi \rightarrow
f\,\bar{f}}\;v_{\textrm{rel}}\,\rangle=
16\,g_A^2\,\alpha\,m_\chi^2\,N_{\textrm{eff}}\,\left(\frac{T}{m_\chi}\right)\,,
\end{equation}
where $N_{\textrm{eff}}$ counts the effective number of kinematically allowed fermionic channels at freeze-out temperature $T_F$.
For each annihilation channel, the contribution to $N_{\textrm{eff}}$ is given by the square of the corresponding fermion charge
multiplied by the color factor whenever applicable.

We can then combine Eqs. (\ref{sigW22}) and (\ref{oldsig}) to obtain the total thermally-averaged annihilation cross section:
\begin{eqnarray}
\langle\,\sigma\,v_{\textrm{rel}}\,\rangle_{\textrm{total}} &=&
g_A^2\,\alpha\,m_\chi^2\, \left[\,16\, N_{\textrm{eff}} + 12\,\left(\,\frac{m_\chi}{m_W} \,\right)^2 \Theta(m_\chi - m_W)\,\right]\,\left(\frac{T}{m_\chi}\right) \,,\nonumber\\
&=& \sigma_0 \left(\frac{T}{m_\chi}\right),
\end{eqnarray}
where $\Theta$ is the Heaviside step function.  Note that annihilation into $W^+W^-$ rapidly becomes
dominant for $m_\chi > m_W$. The annihilation is purely $p$-wave, so we can use the same
expressions as in Ref. \cite{Ho:2012bg} for the relic abundance \cite{ST,KT},
\begin{equation}
\label{Omega}
\Omega_{\chi}\, h^2 = (\,2.14 \times 10^9\,)\, \frac{x_F^{2}\; ({\rm GeV})^{-1}}{g_*^{1/2}\, M_{Pl}\,
\sigma_0},
\end{equation}
with $x_F=m_{\chi}/T_F$ given by
\bea
\label{xf}
x_F &=& \ln \left[\,0.076\, \left(\,\frac{g_\chi}{g_*^{1/2}}\,\right)\, M_{Pl}\, m_\chi\, \sigma_0\,\right]
\nonumber \\
&& ~
- \frac{3}{2}\, \ln\, \ln\, \left[\,0.076\, \left(\,\frac{g_\chi}{g_*^{1/2}}\,\right) \,M_{Pl}\, m_\chi\, \sigma_0\,\right]\,.
\eea
In these equations,
$\Omega_{\chi}$ is the dark matter fraction relative to the critical density,
$h$ is the Hubble parameter in units of 100 km sec$^{-1}$ Mpc$^{-1}$, $g_*$ is the number of
relativistic degrees of freedom in the universe when $\chi$ drops out of thermal equilibrium,
$M_{Pl}$ is the Planck mass, and $g_\chi = 2$ is the number of internal degrees of freedom for the Majorana $\chi \chi$ pair.

We now set $\Omega_\chi h^2$ equal to the latest measurement from the PLANCK experiment~\cite{PLANCK}, $\Omega_{DM} h^2 = 0.12$.
The correct $g_A$ as a function of ADM mass is plotted in Fig.~\ref{fig:bounds}.  It is clear from this figure that an ADM particle with a
thermal relic abundance is ruled out by the LHC for $m_\chi < 100$ GeV.

\section{Discussion}

In this paper, we have studied the collider monojet signals of the anapole dark matter. Our results indicate that the LHC provides useful
constraints on the ADM model, namely, $m_\chi > 100$ GeV to be consistent with thermal relic density and current LHC bounds. It should be
pointed out that like many effective operator scenarios, increasingly high beam energy reach at colliders enters the energy range that is close to the new physics scale $\Lambda$ for even a large coupling constant $g \sim 1$ in $g_A=g/\Lambda^2$.
As shown in Fig.~\ref{fig:bounds}, $g \sim 1$ leads to a minimal CMS allowed $\Lambda$ at half a TeV. With a complete theory, new physics at $\Lambda$ would emerge and yield corrections to the monojet cross-section compared to that from the effective anapole operator.
But the potential correction is highly dependent on the details in the UV theory. The results here should be considered as a qualitative model-independent analysis of the LHC's constraint on the new physics scale $\Lambda$, which gives the anapole coupling at low energy exchanges.

In Ref.~\cite{Ho:2012bg}, it was shown that the differential scattering rate for anapole dark matter at direct detection experiments reaches a maximum around $m_\chi \sim 30-40 $ GeV and it lies just below the threshold for detection by XENON100 \cite{XENON100}. Given the significantly
improved sensitivity around this regime by LUX \cite{LUX}, it may be possible that anapole dark matter with $m_\chi \sim 30-40 $ GeV is ruled out. However, we have just shown that the current LHC results have already excluded anapole dark matter with $m_\chi < 100 $ GeV.
So the new bounds from LUX are redundant for $m_\chi < 100 $ GeV. For $m_\chi > 100 $ GeV, the annihilation channels $\chi\chi\rightarrow W^+W^-$
and $\chi\chi\rightarrow t\bar{t}$ open up and the correct relic abundance is achieved with a much smaller $g_A$. Since the differential scattering rate is proportional to $g_A$, the analysis in Ref.~\cite{Ho:2012bg} indicates that the bound from LUX on anapole dark matter with
$m_\chi > 100 $ GeV is far too loose to exclude this mass range.

Finally, recall from Eq. \eqref{sigW} that $\sigma_{\chi \chi \rightarrow W^{+}W^{-}} \,v_{\textrm{rel}}$ grows quadratically with
$\frac{m_\chi^2}{m_W^2}$ when $m_\chi \gg m_W$. This cross-section may violate the unitarity bound if $m_\chi$ is above $\sim$TeV. With such ADM
masses, it becomes necessary to include additional interactions, e.g., the weak interactions involving the $Z$ bosons. In fact, this is very similar to the protection from a divergent $\sigma_{e^+ e^- \rightarrow W^{+}W^{-}}$ in the standard model. Namely, one will encounter unitarity violation at high energy if one only considers the annihilation channel with the photons but neglects those with the $Z$ bosons. We will explore the quantitative effect of the $Z$ bosons in a future study.

\section{Acknowledgements}
We thank Tao Han, Suichi Kunori, Vikram Rentala and Chien-Peng Yuan for helpful discussions.
C.M.H. and R.J.S. were supported in part by the Department of Energy (DE-FG05-85ER40226 and DE-FG02-96ER40969).

\begin{appendix}

\section{~~~Kinematic Factors}
\label{app:msq}

The kinematic factor ${\cal K}$ in Eq.~(\ref{eq:Melement1}) is
\bea
 {\cal K}&=& \left(\frac{4}{9}\right)\, \bigg\{ 2\left(1-
x_j\right)\, \left[\,4+2\cos^2\theta_\chi q_\chi^2 +2\cos\theta_j\cos\theta_\chi q_\chi
x_j +2x_\chi^2 +2x_\chi x_j  \right. \\ \nn
 && \left. -\,4\,(x_\chi+x_j)+x_j^2(1+\cos^2\theta_j) \,\right]
  - \frac{m_\chi^2}{E_0^2}\,\left[\,4(1-x_j)+(1+\cos^2\theta_j)x_j^2\;\right]\,\bigg\},
\eea
where $x_j = \frac{E_j}{E_0}$,\, $x_\chi = \frac{E_\chi}{E_0}$ and $q_\chi=\frac{|\vec{q}|}{E_0}$ denote the ratios of
final state particle energy and/or 3-momentum to the COM energy $E_0$. In this diagram, $j$ is the radiated gluon and $\chi$ can be either one of the two dark matter particles.

The kinematic factor ${\cal K'}$ in Eq.~(\ref{eq:Melement2}) is
\bea
 {\cal K'}&=&\left(\frac{1}{6}\right)\,\bigg\{\,
 2(1-x_j)\,\left[\,12+2\cos^2\theta_\chi q_\chi^2+10x_\chi^2-12x_j+4\cos\theta_j x_j \right.\\ \nn
 && \left. +\,(5-2\cos\theta_j+\cos^2\theta_j)x_j^2
 -2 x_\chi(10-5x_j+\cos\theta_j x_j)\right. \\ \nn
&& \left. +\,2\cos\theta_\chi q_\chi (2-2x_\chi-x_j+\cos\theta_j x_j)\;\right] \\ \nn
&& -\,\frac{m_\chi^2}{E_0^2}\, \left[\,4-4(1-\cos\theta_j)x_j+(5-2\cos\theta_j+\cos^2\theta_j)x_j^2\;\right]\,\bigg\},
\eea
where $j$ denotes the light quark jet from gluon splitting.

In both ${\cal K}$ and ${\cal K'}$, the color factors are given in the front.
Note that all variables are measured in the COM frame, and the squared amplitudes include diagrams with permuted initial state partons. 

\end{appendix}


\begin{thebibliography}{99}

\bibitem{axion}
L.~D.~Duffy and K.~van Bibber,
New J.\ Phys.\  {\bf 11}, 105008 (2009).

\bibitem{Steigman}
P. Langacker and G. Steigman,
\prd {\bf 84}, 065040 (2011).

\bibitem{Pospelov}
M. Pospelov and T. ter Veldhuis, Phys. Lett. B {\bf 480}, 181 (2000).

\bibitem{Sigurdson}
K. Sigurdson, M. Doran, A. Kurylov, R. R. Caldwell, and M. Kamionkowski,
\prd {\bf 70}, 083501 (2004); erratum, \prd {\bf 73}, 089903 (2006).

\bibitem{Gardner}
S. Gardner, \prd {\bf 79}, 055007 (2009).

\bibitem{Masso}
E. Masso, S. Mohanty, and S. Rao, \prd {\bf 80}, 036009 (2009).

\bibitem{Fitzpatrick}
A. L. Fitzpatrick and K. M. Zurek, \prd {\bf 82}, 075004 (2010).

\bibitem{Cho} W. S. Cho, et al., Phys. Lett. B {\bf 687}, 6 (2010);
erratum, Phys. Lett. B {\bf 694}, 496 (2011).

\bibitem{Heo1} J. H. Heo, Phys. Lett. B {\bf 693}, 255 (2010).

\bibitem{Heo2} J. H. Heo, Phys. Lett. B {\bf 702}, 205 (2011).

\bibitem{Banks}
T. Banks, J.-F. Fortin, and S. Thomas, arXiv:1007.5515 [hep-ph].

\bibitem{Barger1}
V. Barger, W.-Y. Keung, and D. Marfatia, Phys. Lett. B {\bf 696}, 74 (2011).

\bibitem{Fortin}
J.-F. Fortin and T. M. P. Tait, Phys.\ Rev.\ D {\bf 85}, 063506 (2012).

\bibitem{Nobile}
E. Del Nobile, et al., JCAP {\bf 1208}, 010 (2012).

\bibitem{Barger2}
V. Barger, W.-Y. Keung, D. Marfatia, and P.-Y. Tseng, Phys.\ Lett.\ B {\bf 717}, 219 (2012).

\bibitem{Heo3}
J. H. Heo and C. S. Kim, Phys.\ Rev.\ D {\bf 87}, 013007 (2013).

\bibitem{Zeldovich}
Ya. B. Zel'dovich, Sov. Phys. JETP {\bf 6}, 1184 (1958).

\bibitem{Ho:2012bg}
  C.~M.~Ho and R.~J.~Scherrer,
  Phys.\ Lett.\ B {\bf 722}, 341 (2013).

\bibitem{Zurek}
A. L. Fitzpatrick and K. M. Zurek, \prd {\bf 82}, 075004 (2010).

\bibitem{Monojet}
  Y.~Bai, P.~J.~Fox and R.~Harnik,
  JHEP {\bf 1012}, 048 (2010).
  P.~J.~Fox, R.~Harnik, J.~Kopp and Y.~Tsai,
  Phys.\ Rev.\ D {\bf 85}, 056011 (2012).

\bibitem{Mono-photon}
  P.~J.~Fox, R.~Harnik, J.~Kopp and Y.~Tsai,
  Phys.\ Rev.\ D {\bf 84}, 014028 (2011).

\bibitem{Mono-Z}
  L.~M.~Carpenter, A.~Nelson, C.~Shimmin, T.~M.~P.~Tait and D.~Whiteson,
  Phys.\ Rev.\ D {\bf 87}, 074005 (2013).

\bibitem{Mono-Higgs}
  A.~A.~Petrov and W.~Shepherd,
  arXiv:1311.1511 [hep-ph];
  L.~Carpenter, A.~DiFranzo, M.~Mulhearn, C.~Shimmin, S.~Tulin and D.~Whiteson,
  arXiv:1312.2592 [hep-ph].

\bibitem{HeavyMajorana}
  T.~Hapola, M.~Jarvinen, C.~Kouvaris, P.~Panci and J.~Virkajarvi,
  arXiv:1309.6326 [hep-ph].

\bibitem{CTEQ6}
  J.~Pumplin, D.~R.~Stump, J.~Huston, H.~L.~Lai, P.~M.~Nadolsky and W.~K.~Tung,
  JHEP {\bf 0207}, 012 (2002).

\bibitem{cms2013:8TeV}
Cern Report number: CMS-PAS-EXO-12-048,\\ available at
\href{http://cds.cern.ch/record/1525585?ln=en}{http://cds.cern.ch/record/1525585?ln=en}

\bibitem{atlas2013:8TeV}
Cern record number: ATLAS-CONF-2012-147,\\
\href{http://cds.cern.ch/record/1493486}{http://cds.cern.ch/record/1493486}

\bibitem{ST} R. J. Scherrer and M. S. Turner, \prd {\bf 33}, 1585 (1986).

\bibitem{KT} E. W. Kolb and M. S. Turner, {\it The Early Universe}, (New York:
Addison-Wesley, 1990).

\bibitem{PLANCK} P.A.R. Ade, et al., arXiv:1303.5076.

\bibitem{XENON100} E. Aprile, et al., Phys.\ Rev.\ Lett.\  {\bf 109}, 181301 (2012).

\bibitem{LUX} D.~S.~Akerib {\it et al.}  [LUX Collaboration],
  arXiv:1310.8214 [astro-ph.CO].

\end{thebibliography}
\end{document}